\documentclass[%
 reprint,
unsortedaddress,
amsmath,amssymb,
prb,
]{revtex4-1}

\usepackage{graphicx}
\usepackage[colorlinks, bookmarks=false,linkcolor=blue, urlcolor=blue, plainpages=false, pdfpagelabels]{hyperref}
\usepackage[version=3]{mhchem}
\usepackage[table,dvipsnames]{xcolor}
\usepackage{multirow}

\graphicspath{{figures/}}

\renewcommand*{\a}[1]{\hat a_{#1}^{\vphantom{\dagger}}}
\newcommand*{\ad}[1]{\hat a_{#1}^{\dagger}}
\renewcommand*{\c}[1]{\hat{c}_{#1}^{\vphantom{\dagger}}}
\newcommand*{\cd}[1]{\hat{c}_{#1}^{\dagger}}

\newcommand*{\ec}[1]{\left[#1\right]}
\newcommand*{\ep}[1]{\left(#1\right)}
\newcommand*{\ket}[1]{\left|#1\right>}
\newcommand*{\bk}[3]{\left<#1\right|#2\left|#3\right>}
\newcommand*{\mean}[1]{\left<#1\right>}
\newcommand*{\hc}{\mathrm{H.c.}}

\newcommand*{\dsum}[1]{\displaystyle\sum_{#1}}
\newcommand*{\dsumd}[3]{\displaystyle\sum_{#1=#2}^{#3}}

\newcommand*{\dprodd}[3]{\displaystyle\prod_{#1=#2}^{#3}}
\newcommand*{\dbigotimesd}[3]{\displaystyle\bigotimes_{#1=#2}^{#3}}
\newcommand*{\dbigoplusd}[3]{\displaystyle\bigoplus_{#1=#2}^{#3}}

\renewcommand*{\cos}[1]{\,\mathrm{cos}\ep{#1}}
\renewcommand*{\exp}[1]{\,\mathrm{e}^{#1}}

\newcommand*{\SU}[2][{}]{$\mathrm{SU}\hspace{-0.5mm}\ep{#2}_{#1}$}
\newcommand*{\fwf}{\ket{\Psi_{G}^{\mathrm{Fermi}}}}
\newcommand*{\gwf}[2]{\ket{\Psi_{G}^{#1}\ep{#2}}}

\newcommand*{\deltaopt}{\delta_{\mathrm{opt}}}

\newcommand*{\eref}[1]{\textcolor{blue}{Eq.\ref{#1}}}
\newcommand*{\tref}[1]{\textcolor{blue}{TAB.~\ref{#1}}}
\newcommand*{\gref}[1]{\textcolor{blue}{FIG.~\ref{#1}}}

\newcommand*{\etal}{\textit{et~al}}

\begin{document}

\title{Variational Monte-Carlo investigation of \SU{N} Heisenberg chains}

\author{J\'er\^ome Dufour}
\author{Pierre Nataf}
\author{Fr\'ed\'eric Mila}
\affiliation{Institute of Theoretical Physics, \'Ecole Polytechnique
F\'ed\'erale de Lausanne (EPFL), CH-1015 Lausanne, Switzerland }

\date{\today}
\begin{abstract}
Motivated by recent experimental progress in the context of ultra-cold
multi-color fermionic atoms in optical lattices, we have investigated the
properties of the \SU{N} Heisenberg chain with totally antisymmetric
irreducible representations, the effective model of Mott phases with $m<N$
particles per site. These models have been studied for arbitrary $N$ and $m$
with non-abelian bosonization [I.~Affleck, Nuclear Physics B 265, 409 (1986);
305, 582 (1988)], leading to predictions about the nature of the ground state
(gapped or critical) in most but not all cases. Using exact diagonalization and
variational Monte-Carlo based on Gutzwiller projected fermionic wave functions,
we have been able to verify these predictions for a representative number of
cases with $N\leq10$ and $m\leq N/2$, and we have shown that the opening of a
gap is associated to a spontaneous dimerization or trimerization depending on
the value of $m$ and $N$. We have also investigated the marginal cases where
abelian bosonization did not lead to any prediction. In these cases,
variational Monte-Carlo predicts that the ground state is critical with
exponents consistent with conformal field theory.
\end{abstract}

\maketitle

\section{Introduction}

The possibility to load ultra cold fermionic atoms in optical lattices opens
new perspectives in the investigation of lattice models of strongly correlated
systems\cite{wu_2003,gorshkov2010}. When the optical lattice is deep enough,
and when the fermionic atoms have an internal degree of freedom that can take
$N$ values (coming for instance from the nuclear spin in alkaline earths), the
appropriate model takes the form of a generalized Hubbard model
\begin{equation*}
	\hat{H}_{\text{Hub}} = -t \dsum{\langle i,j\rangle}\sum_{\alpha=1}^N (\ad{i\alpha}\a{j\alpha}+\hc)
	+\frac{U}{2}\sum_i \left(\sum_\alpha \hat n_{i\alpha}\right)^2
	\label{eq:SUN-Hubbard}
\end{equation*}
where $\ad{i\alpha}$ and $\a{i\alpha}$ are creation and annihilation fermionic
operators, $\hat n_{i\alpha}=\ad{i\alpha}\a{i\alpha}$, $t$ is the hopping
integral between pairs of nearest neighbors $\langle i,j\rangle$, and $U$ is
the on-site repulsion.

When the average number of atoms per site $m$ is an integer, and for large
enough $U/t$, the system is expected to be in a Mott insulating phase
\cite{assaraf1999}. Fluctuations induced by the hopping term in the manifold of
states with $m$ fermions per site start at second order in $t/U$, and the
processes that appear at this order consist in exchanging particles between
pairs of neighboring sites, leading to the effective Hamiltonian
\begin{equation}
	\hat{H}_{\text {eff}}= \frac{2 t^2}{U}\hat{H}
	\label{eq:SUN-effective}
\end{equation}
with 
\begin{equation}
	\hat{H} = \dsum{\langle i,j\rangle}\sum_{\alpha,\beta=1}^N\ad{i\alpha}\a{i\beta}\ad{j\beta}\a{j\alpha}
	\label{eq:SUN-heisenberg}
\end{equation}
In the case of electrons with spin $\uparrow$ or $\downarrow$, this Hamiltonian
has \SU{2} symmetry, and it is equivalent to the Heisenberg model with coupling
constant $4t^2/U$ thanks to the identity
\begin{equation*}
	\sum_{\alpha,\beta=\uparrow,\downarrow}\ad{i\alpha}\a{i\beta}\ad{j\beta}\a{j\alpha}=2 \vec{S}_i\cdot\vec{S}_j+\frac{1}{2}\hat n_i \hat n_j
\end{equation*}
and to the fact that $\hat n_i \hat n_j$ is a constant in the manifold of
states with one particle per site.

More generally, when the number of degrees of freedom is equal to $N>2$, Mott
phases can be realized for all integer values of $m<N$. The effective model now
has \SU{N} symmetry. This can be made explicit by introducing the generators 
\begin{equation*}
	\hat{S}_{\alpha\beta} = \ad{\alpha}\a{\beta} - \dfrac{m}{N}\delta_{\alpha\beta}
\end{equation*}
which satisfy the \SU{N} algebra
\begin{equation*}
	\ec{\hat{S}_{\alpha\beta},\hat{S}_{\mu\nu}} = \delta_{\mu\beta}\hat{S}_{\alpha\nu}-\delta_{\alpha\nu}\hat{S}_{\mu\beta}
\end{equation*}
thanks to the identity
\begin{equation*}
	\sum_{\alpha,\beta=1}^N\hat{S}^{i}_{\alpha\beta}\hat{S}^{j}_{\beta\alpha}
	= \sum_{\alpha,\beta=1}^N\ad{i\alpha}\a{i\beta}\ad{j\beta}\a{j\alpha} - \dfrac{m^{2}}{N}
\end{equation*} 
In the \SU{N} language, working with $m$ fermions per site corresponds to
working with the totally antisymmetric irreducible representation (irrep) that
can be represented by a Young tableau with $m$ boxes in one column. For any
allowed $m$, there is a conjugate equivalent representation: a system with
$m=k$ particles per site is equivalent to a system with $m=N-k$ particles per
site.

The model \eref{eq:SUN-heisenberg} captures the low-energy physics of
multi-color ultra-cold atoms in optical lattices, systems for which remarkable
progress has been recently achieved on the experimental side. For instance, the
\SU{N}-symmetry has been observed in ultracold quantum gas of fermionic
\ce{^{128}Yb}~\cite{Scazza2014} or \ce{^{87}Sr}~\cite{Zhang2014}. Another
example~\cite{Pagano2014} is the realization of one dimensional quantum wires
of repulsive fermions with a tunable number of components.

The \SU{N} Heisenberg model with the fundamental representation at each site
($m=1$), which corresponds to the Mott phase with one particle per site, has
been investigated in considerable detail over the years. In one dimension,
there is a Bethe ansatz solution for all values of $N$~\cite{Sutherland1975},
and Quantum Monte Carlo simulations free of minus sign problem have given
access to the temperature dependence of correlation
functions~\cite{Frischmuth1999, Messio2012}. In two dimensions, a number of
simple lattices have been investigated for a few values of $N$ with a
combination of semiclassical, variational and numerical approaches, leading to
a number of interesting predictions at zero temperature~\cite{li1998,
vandenbossche2000, vandenbossche2001, mambrini2003, arovas2008, wang_z2_2009,
toth2010, wu2011, szirmai2011, corbozSU42011, szirmaiSU62011,
corbozsimplex2012, corbozPRX2012, bauer2012, corboz-2013,wu_2014}.

In comparison, the case of higher antisymmetric irreps ($m>1$) has been little
investigated. In 2D, there is a mean-field prediction that chiral phases might
appear for large $m$ provided $N/m$ is large enough~\cite{hermele2009,
hermele_topological_2011}, and some cases of self-conjugate irreps such as the
6-dimensional irrep of \SU{4} have been investigated with Quantum Monte Carlo
simulations~\cite{assaad2005, cai2013, lang2013, zhou2014} and variational
Monte Carlo\cite{paramekanti_2007}.  In 1D, apart from a few specific
cases~\cite{andrei1984, johannesson1986, paramekanti_2007,fuhringer2008,
rachel2009, nonne2011, nonne2013, morimoto2014, quella2012}, including more
general irreps than simply the totally antisymmetric ones, the most general
results have been obtained by Affleck quite some time ago~\cite{Affleck1986a,
Affleck1988}.  Applying non-abelian bosonization to the weak coupling limit of
the \SU{N} Hubbard model, he identified two types of operators that could open
a gap: Umklapp terms if $N>2$ and $N/m=2$, and higher-order operators with
scaling dimension $\chi=N(m-1)m^{-2}$ allowed by the $\mathbb{Z}_{N/m}$
symmetry if $N/m$ is an integer strictly larger than 2. This allowed him to
make predictions in four cases: i) $N/m$ is not an integer: the system should
be gapless because there is no relevant operator that could open a gap; ii)
$N>2$ and $N/m=2$: the system should be gapped because Umklapp terms are
always relevant; iii) $N/m$ is an integer strictly larger than 2 and $\chi<2$:
the system should be gapped because there is a relevant operator allowed by
symmetry. This case is only realized for \SU{6} with $m=2$; iv) $N/m$ is an
integer strictly larger than 2 and $\chi>2$: the system should be gapless
because there is no relevant operator allowed by symmetry. The only case where
the renormalization group argument based on the scaling dimension of the
operator does not lead to any prediction is the marginal case $\chi=2$, which
is realized for two pairs of parameters: (\SU{8} $m=2$) and (\SU{9} $m=3$). These
predictions are summarized in \tref{rlt-N-vs-m-gap}. Finally, in all gapless
cases, the system is expected to be in the \SU[k=1]{N} Wess-Zumino-Witten
universality class~\cite{Knizhnik1984, Affleck1986a}, with algebraic
correlations that decay at long distance with a critical exponent $\eta=2-2/N$.

\begin{table}[h]
	\centering
	\setlength{\tabcolsep}{-0.1pt}
	\begin{tabular}{||l|p{4mm}p{4mm}p{4mm}p{4mm}p{4mm}p{4mm}p{4mm}p{4mm}||c|l}
		\cline{1-10}
		\hspace{4mm}$N=$\hspace{1mm} & \hspace{1mm}3 & \hspace{1mm}4 & \hspace{1mm}5 & \hspace{1mm}6 & \hspace{1mm}7 & \hspace{1mm}8 & \hspace{1mm}9 & 10&\cellcolor{Gray}$N/m\notin\mathbb{N}$&\multirow{3}{*}{$\left.\phantom{\rule{0.1cm}{0.65cm}}\right\}$ Gapless}\\
		\cline{1-1}
		\hspace{1mm}$m=1$&\cellcolor{CadetBlue}&\cellcolor{CadetBlue}&\cellcolor{CadetBlue}&\cellcolor{CadetBlue}&\cellcolor{CadetBlue}&\cellcolor{CadetBlue}&\cellcolor{CadetBlue}&\cellcolor{CadetBlue}&\cellcolor{CadetBlue} $m=1$\\
		\hspace{1mm}$m=2$&\cellcolor{lightgray}&\cellcolor{LimeGreen}&\cellcolor{Gray}&\cellcolor{yellow}&\cellcolor{Gray}&\cellcolor{orange}&\cellcolor{Gray}&\cellcolor{CarnationPink}&\cellcolor{CarnationPink} $\chi>2$\\
		\hspace{1mm}$m=3$&&\cellcolor{lightgray}&\cellcolor{lightgray}&\cellcolor{LimeGreen}&\cellcolor{Gray}&\cellcolor{Gray}&\cellcolor{orange}&\cellcolor{Gray}& \cellcolor{orange}$\chi=2$& \hspace{10mm}?\\
		\hspace{1mm}$m=4$&&&\cellcolor{lightgray}&\cellcolor{lightgray}&\cellcolor{lightgray}&\cellcolor{LimeGreen}&\cellcolor{Gray}&\cellcolor{Gray}&\cellcolor{yellow}$\chi<2$&\multirow{2}{*}{$\left.\phantom{\rule{0.1cm}{0.5cm}}\right\}$ Gapped}\\
		\hspace{1mm}$m=5$&&&&\cellcolor{lightgray}&\cellcolor{lightgray}&\cellcolor{lightgray}&\cellcolor{lightgray}&\cellcolor{LimeGreen}&\cellcolor{LimeGreen} $N/m=2\;$\\
		\cline{1-10}
	\end{tabular}
	\caption{
	Summary of the predictions of Refs.~\onlinecite{Affleck1986a,Affleck1988}
	for a representative range of \SU{N} with $m$ particles per site. Note
	that models with $m=k$ and $m=N-k$ are equivalent up to a constant.
	Therefore the light gray shaded region can be deduced from the other cases
	and does not need to be studied.
	}
	\label{rlt-N-vs-m-gap}
	\setlength{\tabcolsep}{6pt}
\end{table}

To make progress on the general problem of the \SU{N} Heisenberg model with
higher antisymmetric irreps, it would be very useful to have flexible yet
reliable numerical methods that would allow to test these predictions in a
systematic way. In particular, the methods should not be limited to 1D, or to
cases such as self-conjugate irreps, for which there is a minus-sign free
Quantum Monte Carlo algorithm. In this paper, we have decided to test the
potential of Gutzwiller projected wave functions by a systematic investigation
of the 1D case discussed by Affleck using variational Monte Carlo (VMC). There
are two main reasons for this choice. First of all, these wave functions have
been shown to be remarkably accurate in the case of the \SU{4} Heisenberg chain
with the fundamental representation by Wang and Viswanath~\cite{wang_z2_2009},
who have in particular shown that they lead to the exact critical exponent in
that case. Besides, this approach can be easily generalized to higher
dimensions, as already shown for the fundamental representation in a number of
cases~\cite{lajko_tetramerization_2013,corboz-2013}. Moreover, it has been used
by Paramekanti and Marston~\cite{paramekanti_2007} for the self-conjugate
representation in one and two dimensions.

In parallel, exact diagonalizations based on the extension of a recent
formulation by two of the present authors~\cite{nataf2014} will be used
whenever possible to benchmark the VMC approach on small clusters and, in some
cases, to actually confirm the physics on the basis of a finite-size analysis.

As we shall see, the combination of these approaches leads to results that
agree with Affleck's predictions whenever available, and to the identification
of the symmetry breaking pattern in the gapped phases. In addition, it predicts
that the marginal cases are gapless with algebraic correlations.

The paper is organized as follows. The next section describes the methods, with
emphasis on the variational wave functions that will be used throughout. The
third section is devoted to a comparison of the results obtained using the
simplest wave functions (with no symmetry breaking) with those of the Bethe
ansatz solution for the $m=1$ case, with the conclusion that the agreement is
truly remarkable. The fourth section deals with the cases where Umklapp
processes are present ($N>2$, $N/m=2$), while the fifth one deals with the
case where there is no Umklapp process but a relevant operator (\SU{6} $m=2$).
The marginal cases are dealt with in the sixth section, and the case where
$N/m$ is an integer without relevant nor Umklapp operators in the seventh
section.  Finally, the critical exponents are computed and compared to
theoretical values for all gapless systems in the eighth section.

\section{The methods}
\subsection{Gutzwiller projected wave functions}

The variational wave functions investigated in the present paper are obtained
from fermionic wave functions that have on average $m$ particles per site by
applying a Gutzwiller projector $\hat{P}_{G}^{m}$ that removes all states with
a number of particles per site different from $m$:
\begin{equation}
	\hat{P}_{G}^{m}=\dprodd{i}{1}{n} \prod_{p\neq m} \frac{\hat n_i-p}{m-p}
\end{equation}
where $n$ is the number of sites, and where the product over $p$ runs over all
values from $0$ to $N$ except $p=m$.

In the present paper, we will concentrate on simple fermionic wave-functions
that, before projection, correspond to the ground state of trial Hamiltonians
that contain only hopping terms. For \SU{2}, the inclusion of pairing terms
have been shown to lead to significant improvements\cite{yunoki}, but the
generalization to \SU{N} is not obvious because one cannot make an \SU{N}
singlet with two sites as soon as $N>2$. In addition, in the case of the
fundamental representation where Bethe ansatz results are available for
comparison, these simple wave functions turn out to lead to extremely precise
results as soon as $N>2$. 

In practice, the construction of a Gutzwiller projected wave function starts
with the creation of a trial Hamiltonian $\hat{T}$ that acts on $n$ sites and
is written with fermionic operators $\a{i\alpha}$ and $\ad{i\alpha}$. When
different colors are involved in $\hat{T}$, and as long as there is no term
mixing different colors, the Hamiltonian can be rewritten as a direct sum: $
\hat{T} = \bigoplus_{\alpha=1}^{N}\hat{T}_{\alpha}$. Then, for each color,
there will be one corresponding unitary matrix $U^{\alpha}$ that diagonalizes
$\hat{T}_{\alpha}$. So the new fermionic operators are given by:
\begin{align*}
	\c{i\alpha} &= \dsumd{j}{1}{n}U_{ij}^{\alpha\dagger}\a{j\alpha}
	& \cd{i\alpha} &= \dsumd{j}{1}{n}U_{ji}^{\alpha}\ad{j\alpha},
\end{align*}
and the trial Hamiltonian can be written in a diagonal basis:
\begin{equation*}
	\hat{T} = \dbigoplusd{\alpha}{1}{N}\dsumd{i}{1}{n}\omega_{i\alpha}\cd{i\alpha}\c{i\alpha}
\end{equation*}
with $\omega_{i\alpha}<\omega_{i+1\alpha}$.

In the Mott insulating phase, the system possesses $nm/N$ particles of each
color and exactly $m$ particles per site. By filling the system with the $nm/N$
lowest energy states of each color, the resulting fermionic wave function
contains $nm$ particles:
\begin{equation}
	\ket{\Psi} = \dbigotimesd{\alpha}{1}{N}\dprodd{i}{1}{nm/N}\cd{i\alpha}\ket{0}
	= \dbigotimesd{\alpha}{1}{N}\dprodd{i}{1}{mn/N}\dsumd{j}{1}{n}U_{ji}^{\alpha}\ad{j\alpha}\ket{0}
\end{equation}
in terms of which the variational wave function is given by
\begin{equation}
	\ket{\Psi_G}=\hat{P}_{G}^{m}\ket{\Psi}.
\end{equation}

Since the Heisenberg model exchanges particles on neighboring sites, the
simplest trial Hamiltonian that allows the hopping of particles and its
corresponding Gutzwiller projected wave function are:
\begin{equation*}
	\hat{T}_{\alpha}^{\mathrm{Fermi}} = \dsumd{i}{1}{n}\ep{\ad{i\alpha}\a{i+1\alpha}+\hc}
	\quad\rightarrow\quad \fwf.
\end{equation*}
In cases where a relevant or Umklapp operator is present, the ground state is
expected to be a singlet separated from the first excited state by a gap, and
to undergo a symmetry breaking that leads to a unit cell that can accommodate a
singlet. In practice, this means unit cells with $d=N/m$ sites. To test for
possible instabilities, we have thus used wave functions that are ground
states of Hamiltonians that creates $d$-merization:
\begin{equation*}
	\hat{T}_{\alpha}^{t_{i}} = \dsumd{i}{1}{n}\ep{t_{i}\ad{i\alpha}\a{i+1\alpha}+\hc}
	\quad\rightarrow\quad \gwf{d}{\delta}.
\end{equation*}
Assuming that the mirror symmetry is preserved, the wave functions
$\gwf{d}{\delta}$ for dimerization ($d=2$) and trimerization ($d=3$) have only
one allowed free parameter $\delta$, and the hopping amplitudes in a unit cell
are given by: 
\begin{equation*}
	\begin{cases}
		t_{i} = 1-\delta &\text{if } i = d\\
		t_{i} = 1 &\text{otherwise.}
	\end{cases}
\end{equation*}
To test for a possible tetramerization for \SU{8} $m=2$, since the unit cell
contains four sites, one additional free parameter is allowed (still assuming
that the mirror symmetry is preserved in the ground state). Therefore, we have
used the wavefunction $\gwf{4}{\delta_{1},\delta_{2}}$ with hopping amplitudes
defined by:
\begin{equation*}
	\begin{cases}
		t_{i} = 1-\delta_{1} &\text{if } i = 2\\
		t_{i} = 1-\delta_{2} &\text{if } i = 4\\
		t_{i} = 1 &\text{otherwise.}
	\end{cases}
\end{equation*}
This method is always well defined for periodic boundary conditions when $N/m$
is even. But when $N/m$ is odd, the ground state is degenerate for periodic
boundary conditions if the translation symmetry is not explicitly broken, and
one has to use anti-periodic boundary conditions for $\fwf$,
$\gwf{d}{0}(d=2,3)$ and $\gwf{4}{0,0}$.

The hope is that if $\hat{T}$ is wisely chosen, then $\ket{\Psi_{G}}$ captures
correctly the physics of the ground state, i.e. with a good variational wave
function, $E_{G}\equiv\bk{\Psi_{G}}{\hat{H}}{\Psi_{G}}\approx E_{0}$, the exact
ground state energy. To check the pertinence of this statement, we have
compared the energies and nearest-neighbor correlations with those computed
with ED on small systems with open boundary conditions. In the table
\tref{rlt-ed-vmc-energies}, one can see, for some systems, the comparison
between ED and VMC results for the ground state energy. The nearest-neighbor
correlations will be compared in the next sections. Considering the excellent
agreement between the two methods for the cluster sizes available to ED, there
are good reasons to hope that these Gutzwiller projected wave functions can
quantitatively describe the properties of the ground state.

\begin{table}[h]
	\centering
	\begin{tabular}{|c|c|c|c|c|c|}
		\hline
		$N$ & $m$ & $n$ & ED & VMC & error [\%] \\\hline
		4 &2&16&-1.6971&-1.6916&-0.33\\
		4 &2&18&-1.6925&-1.6866&-0.35\\
		6 &2&15&-2.7351&-2.7287&-0.23\\
		6 &3&12&-4.0295&-4.0261&-0.08\\
		6 &3&14&-4.0162&-4.0123&-0.10\\
		8 &2&12&-3.1609&-3.1587&-0.07\\
		8 &2&16&-3.1857&-3.1828&-0.09\\
		9 &3&9 &-6.0960&-6.0810&-0.25\\
		9 &3&12&-6.1162&-6.0980&-0.30\\
		10&2&15&-3.3992&-3.3919&-0.21\\
		\hline
	\end{tabular}
	\caption{
	Comparison between the ED and VMC energies per site. The incertitudes on
	the VMC data are smaller than $10^{-4}$. The relative error is always
	smaller than $0.35\%$.
	}
	\label{rlt-ed-vmc-energies}
\end{table}

\subsection{Exact diagonalizations}

On a given cluster, the total Hilbert space grows very fast with $N$, and the
standard approach that only takes advantage of the conservation of the color
number is limited to very small clusters for large $N$. Quite recently, two of
the present authors have developed a simple method to work directly in a given
irrep for the \SU{N} Heisenberg model with the fundamental representation at
each site~\cite{nataf2014}, allowing to reach cluster sizes typical of \SU{2}
for any $N$. This method can be extended to the case of more complicated
irreps at each site, in particular totally antisymmetric irreps, and the exact
diagonalization results reported in this manuscript have been obtained along
these lines.

\subsection{Correlation function and structure factor}\label{rls-cor-sf}
To characterize the ground state, it will prove useful to study the diagonal
correlation defined by:
\begin{equation}\label{rle-lrc}
	C\ep{r} = \dsum{\alpha}\mean{\hat{S}_{\alpha\alpha}^{0}\hat{S}_{\alpha\alpha}^{r}}
	= \dsum{\alpha}\mean{\ad{0\alpha}\a{0\alpha}\ad{r\alpha}\a{r\alpha}} - \dfrac{m^{2}}{N}.
\end{equation}
The structure factor is then given by the Fourier transform of this function:
\begin{equation}\label{rle-sf}
	\tilde{C}\ep{k}= \dfrac{1}{2\pi}\dfrac{N}{m\ep{N-m}} \dsum{r}C\ep{r}\exp{ikr}
\end{equation}
where the prefactor has been chosen such that:
\begin{equation*}
	\dsum{k}\tilde{C}\ep{k} = \dfrac{n}{2\pi}.
\end{equation*}

\section{\SU{N} with $m=1$}\label{rls-m1}
In this section, we extend the \SU{4} results of Wang and
Vishwanath~\cite{wang_z2_2009} to arbitrary $N$ for $m=1$ (fundamental
representation), and we perform a systematic comparison with Bethe ansatz and
Quantum Monte-Carlo (QMC) results. Since these systems are known to be gapless,
$\fwf$ is the only relevant wave function to study.

\begin{figure}[h]
	\begin{center}
		\includegraphics[scale=0.7]{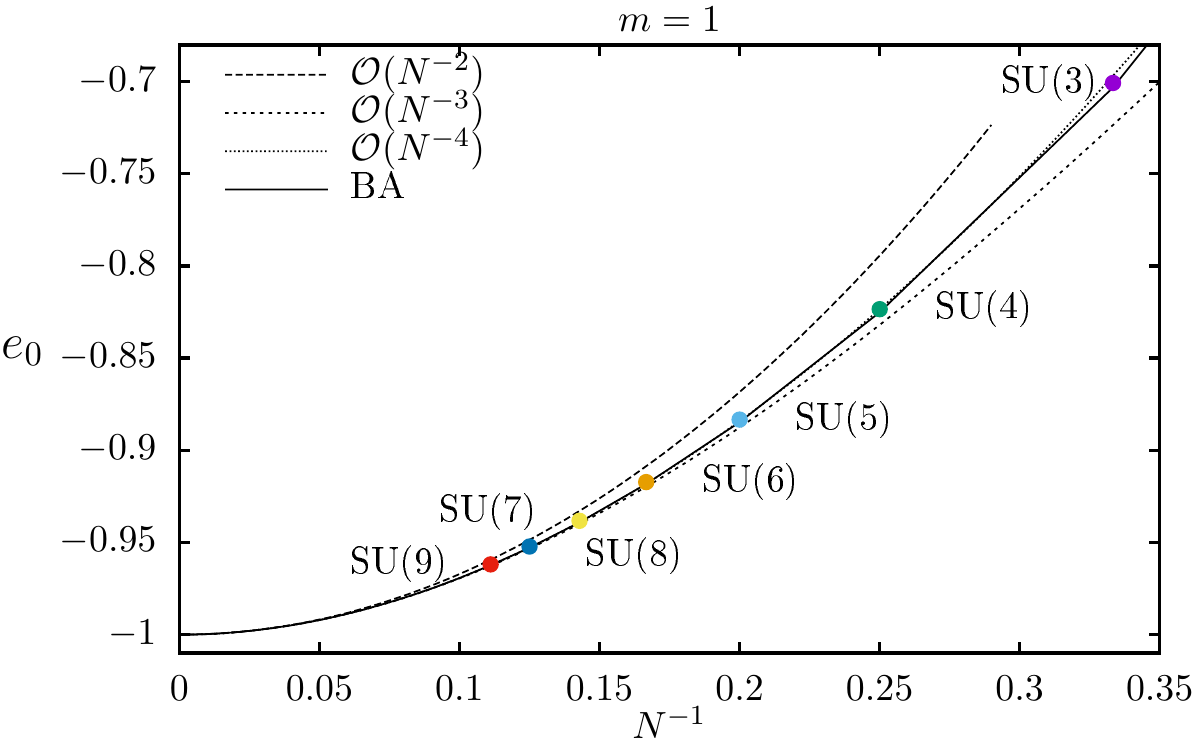}
	\end{center}
	\vspace{-0.6cm}
	\caption{
	Variational energy per site of \SU{N} chains with the fundamental irrep at
	each site (dots) compared to Bethe ansatz exact results (solid line) and
	polynomial approximations in $1/N$ (dashed lines).
	}
	\label{rlg-sun-m1-vmc-ba}
\end{figure}

Let us start with the ground state energy. Using Bethe ansatz,
Sutherland~\cite{Sutherland1975} derived an exact formula for the ground state
energy per site $e_0(N)$ of the Hamiltonian \eref{eq:SUN-heisenberg} that
can be written as a series in powers of $1/N$:
\begin{equation}
	e_0(N)=-1+2 \dsumd{k}{2}{\infty} \frac{(-1)^k \zeta(k)}{N^k}
\end{equation}
where $\zeta(k)=\sum_{n=1}^\infty (1/n^k)$ is Riemann's zeta function.
$e_0(N)$ is depicted in \gref{rlg-sun-m1-vmc-ba} as a continuous line. The
dashed lines are approximations obtained by truncating the exact solution at
order $N^{-k},\,k\geq 2$. For comparison, the variational energies obtained in
the thermodynamic limit after extrapolation from finite size systems are shown
as dots in \gref{rlg-sun-m1-vmc-ba}. The agreement with the exact solution is
excellent for all values of $N$, and it improves when $N$ increases (see table
\tref{rlt-m1}). Quite remarkably, the variational estimate is better than the
$N^{-4}$ estimate even for \SU{3}.

\begin{table}[h]
	\centering
	\begin{tabular}{|c|c|c|c|}
		\hline
		$N$& BA & VMC& error [\%]\\ \hline
		3& -0.7032& -0.7007& -0.36\\
		4& -0.8251& -0.8234& -0.21\\
		5& -0.8847& -0.8833& -0.16\\
		6& -0.9183& -0.9173& -0.11\\
		7& -0.9391& -0.9383& -0.09\\
		8& -0.9528& -0.9522& -0.06\\
		9& -0.9624& -0.9620& -0.05\\
		\hline
	\end{tabular}
	\caption{
	Comparison of the variational energies for $m=1$ systems obtained for
	infinite chains with exact Bethe ansatz. The incertitudes on the VMC data
	are smaller than $10^{-4}$.
	}
	\label{rlt-m1}
\end{table}

\begin{figure}[t]
	\begin{center}
		\includegraphics[scale=0.7]{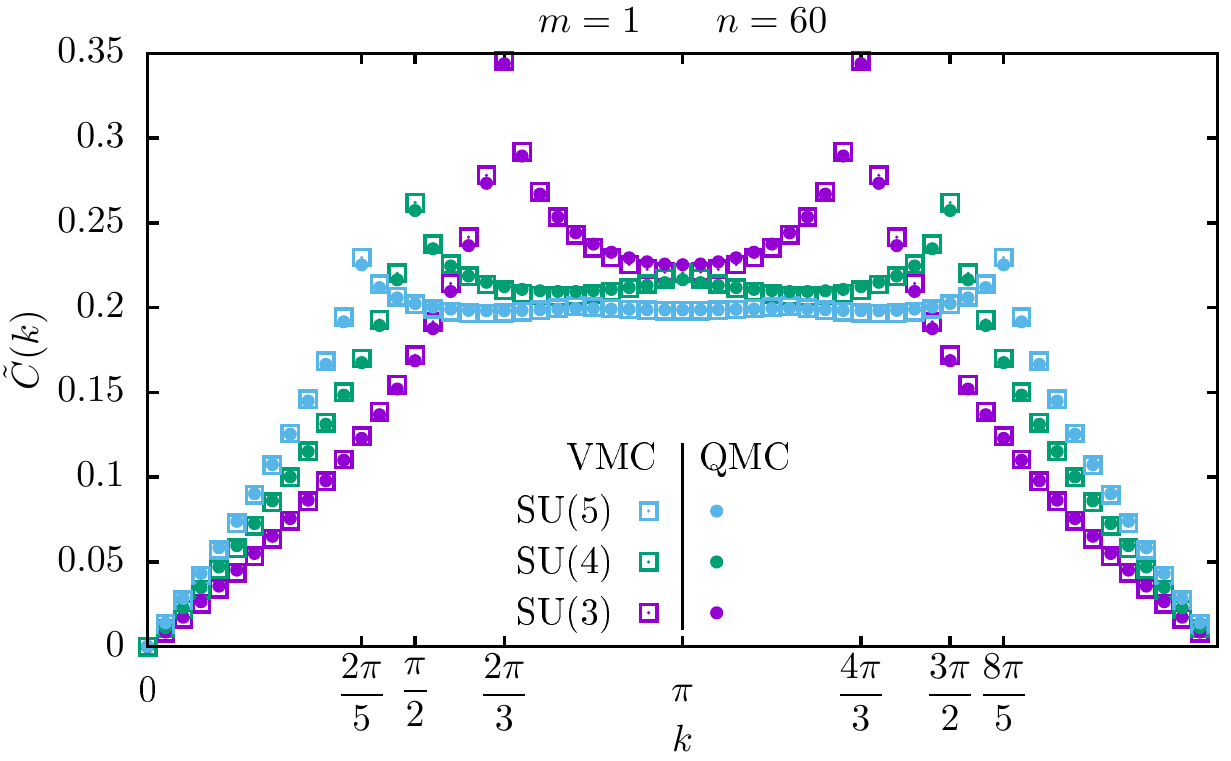}
	\end{center}
	\vspace{-0.6cm}
	\caption{
	Comparison of the structure factors calculated with VMC (empty squares)
	and QMC (filled circles) for various \SU{N} systems. In the VMC
	calculations, anti-periodic boundary conditions have been used for \SU{3}
	and \SU{5}, and periodic ones for \SU{4}.
	}
	\label{rlg-qmc-vmc-lrc}
\end{figure}

We now turn to the diagonal correlations and its associated structure factor
defined by \eref{rle-lrc} and \eref{rle-sf}. At very low
temperature, QMC has been used by Frischmuth \etal~\cite{Frischmuth1999} for
\SU{4} and by Messio and Mila~\cite{Messio2012} for various values of $N$ to
compute this structure factor. The QMC data of Messio and Mila and the results
obtained with VMC for $n=60$ sites are shown in \gref{rlg-qmc-vmc-lrc}.
Qualitatively, the agreement is perfect: VMC reproduces the singularities
typical of algebraically decaying long-range correlations. But even
quantitatively the agreement is truly remarkable, and, as for the ground state
energy, it improves when $N$ increases. Clearly, Gutzwiller projected wave
functions capture the physics of the $m=1$ case very well.

\section{\SU{N} with $m=N/2$}\label{rls-Nm2}
For these systems, there is a self-conjugate antisymmetric representation of
\SU{N} at each site. The ground states of such systems, referred to as
extended valence bound solids~\cite{Affleck1991}, are predicted to break the
translational symmetry, to be two-fold degenerate and to exhibit dimerization
since only two sites are needed to create a singlet, and the spectrum is
expected to be gapped.

\begin{figure}[h!]
	\includegraphics[scale=0.7]{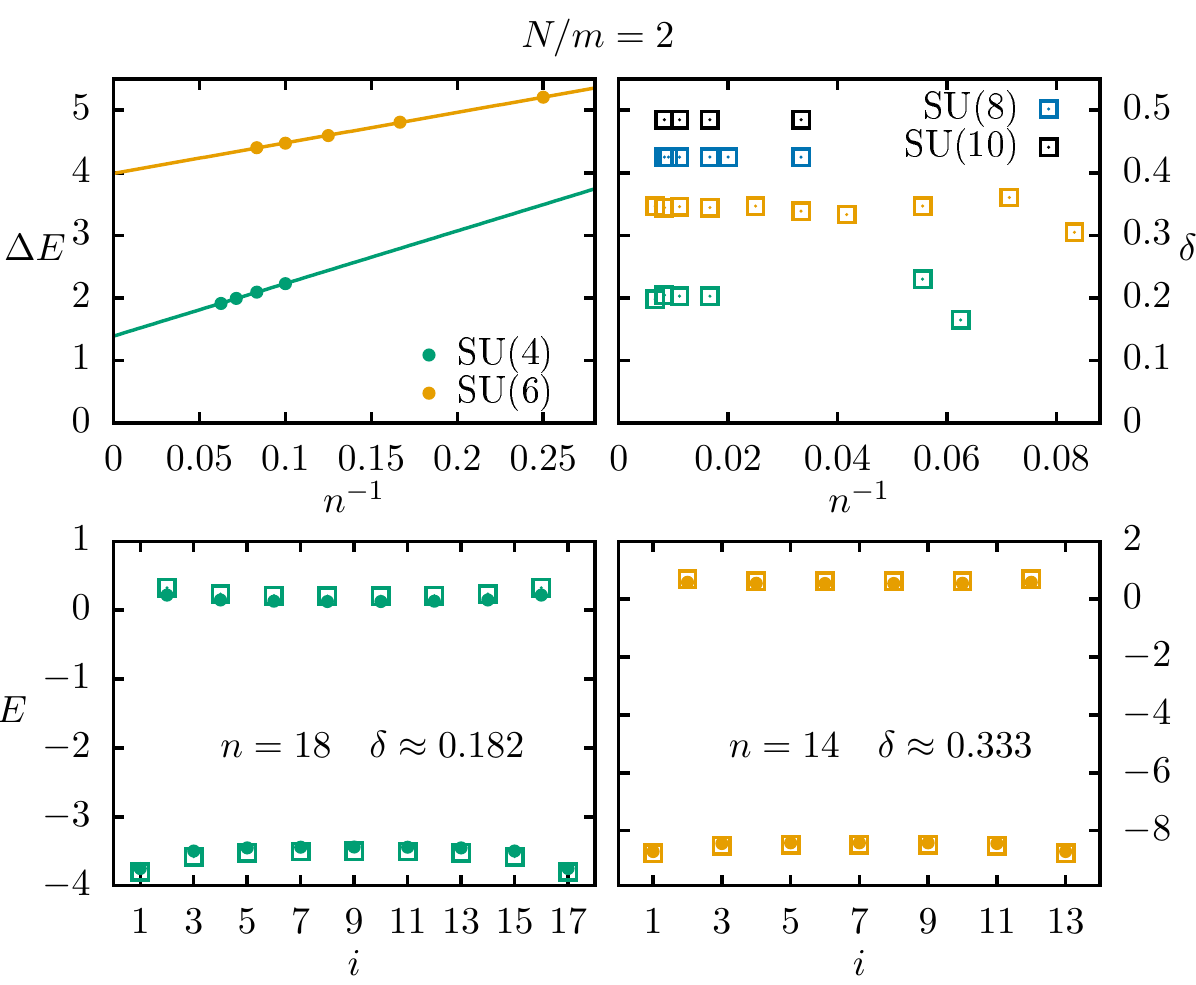}
	\vspace{-0.2cm}
	\caption{
	ED and VMC results for various \SU{N} models with $m=N/2$. Upper left
	panel: size dependence of the energy gap for \SU{4} and \SU{6}. Upper
	right panel: optimal variational parameter $\delta$ for periodic boundary
	conditions for \SU{4}, \SU{6}, \SU{8}, and \SU{10}. Lower panels: energy
	per bond for \SU{4} (left) and \SU{6} (right) calculated with ED (circles)
	and VMC (squares) for open boundary conditions. Note that the optimal
	variational parameters $\deltaopt$ are different in the upper right panel
	and in the lower panels because they correspond to different boundary
	conditions (periodic and open).
	}
	\label{rlg-Nom2-gap-delta-nnc}
\end{figure}

We have investigated two representative cases, (\SU{4} $m=2$) and (\SU{6} $m=3$),
with ED up to 18 and 14 sites respectively, and the cases $N=4$ to $10$ with
VMC. The main results are summarized in \gref{rlg-Nom2-gap-delta-nnc}.

Let us start by discussing the ED results. Clusters with open boundary
conditions have been used because they are technically simpler to handle with
the method of Ref.~\onlinecite{nataf2014}, and because, in the case of
spontaneous dimerization, they give directly access to one of the broken
symmetry ground states if the number of sites is even. The gap as a function of
the inverse size is plotted in the upper left panel of
\gref{rlg-Nom2-gap-delta-nnc} for \SU{4} and \SU{6}. In both cases, the results
scale very smoothly, and a linear fit is consistent with a finite and large
value of the gap in the thermodynamic limit. In the lower panels of
\gref{rlg-Nom2-gap-delta-nnc}, the bond energy is plotted as a function of the
bond position for the largest available clusters (18 sites for \SU{4}, 14 sites
for \SU{6}) with solid symbols. A very strong alternation between a strongly
negative value and an almost vanishing (slightly positive) value with very
little dependence on the bond position clearly demonstrates that the systems
are indeed spontaneously dimerized.

Let us now turn to the the VMC results. Since the relevant instability is a
spontaneous dimerization, it is expected that the dimerized $\gwf{2}{\delta}$
wave function allows one to reach lower energy than the $\fwf$ one. This is
indeed true for all cases we have investigated (up to $N=10$ and to $n\gtrsim
100$), and the optimal value of the dimerization parameter $\deltaopt>0$ is
nearly size independent and increases with $N$ (see upper right panel of
\gref{rlg-Nom2-gap-delta-nnc}), in qualitative agreement with the gap increase
between \SU{4} and \SU{6} observed in ED. To further benchmark the Gutzwiller
projected wave functions for these cases, we have calculated the bond energy
using the optimal value of $\delta$ (open symbols in the lower panel of
\gref{rlg-Nom2-gap-delta-nnc}) for the same clusters as those used for ED with
open boundary conditions. The results are in very good quantitative agreement.

\begin{figure}[h]
	\begin{center}
		\includegraphics[scale=0.7]{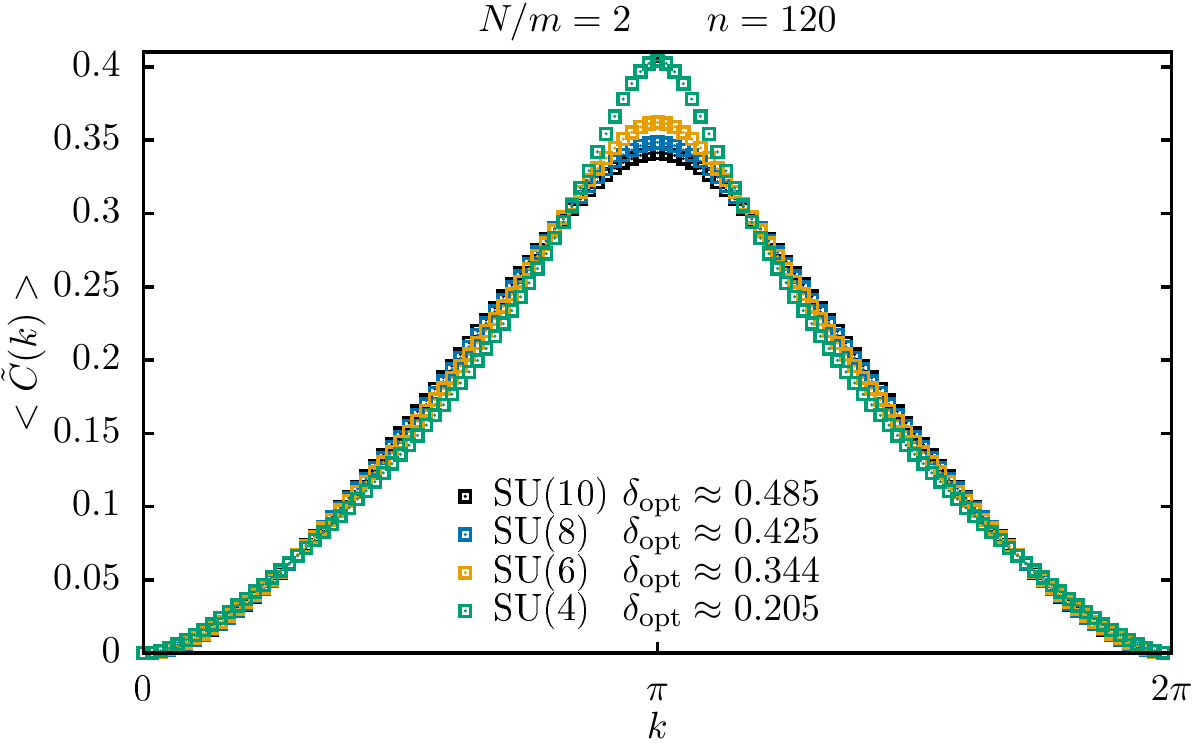}
	\end{center}
	\vspace{-0.6cm}
	\caption{
	Structure factor of various \SU{N} models with $m=N/2$ calculated with VMC
	with the optimal variational parameter $\deltaopt$.
	}
	\label{rlg-Nom2-sf}
	\vspace{-0.5cm}
\end{figure}

With the large sizes accessible with VMC, it is also interesting to calculate
the diagonal structure factor defined in \eref{rle-sf}. All the structure
factors peak at $k=\pi$, but, unlike in the case of the fundamental
representation, there is no singularity but a smooth maximum (see
\gref{rlg-Nom2-sf}). This shows that the antiferromagnetic correlations
revealed by the peak at $k=\pi$ are only short ranged, and that the
correlations decay exponentially at long distance, in agreement with the
presence of a gap, and with the spontaneous dimerization.

To summarize, ED and VMC results clearly support Affleck's predictions that the
$N/m=2$ systems are gapped and point to a very strong spontaneous dimerization
in agreement with previous results by Paramekanti and
Marston~\cite{paramekanti_2007}.

\section{\SU{6} with $m=2$}
\begin{figure}[h]
	\begin{center}
		\includegraphics[scale=0.7]{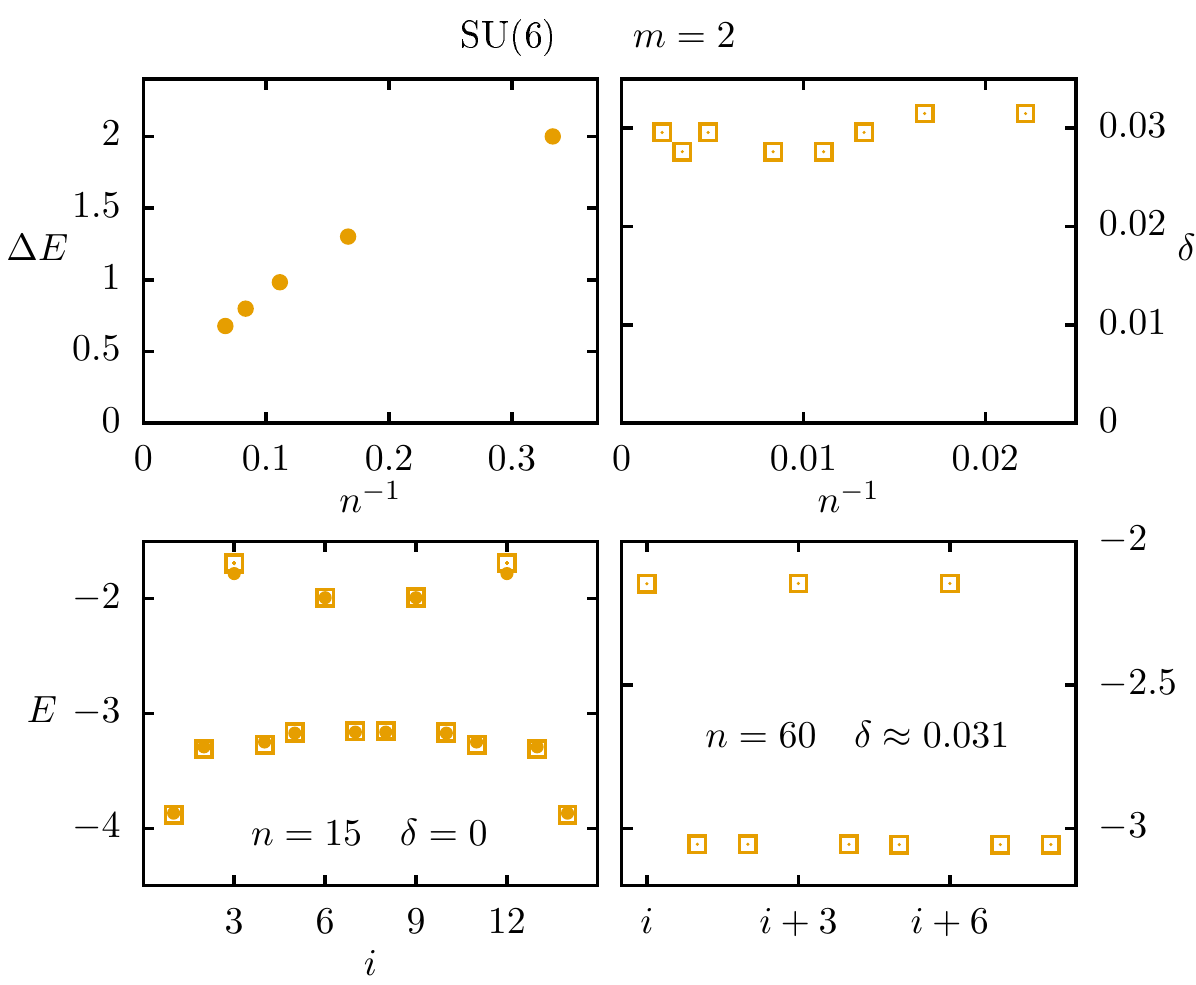}
	\end{center}
	\vspace{-0.5cm}
	\caption{
	ED and VMC results for the \SU{6} model with $m=2$. Upper left panel: size
	dependence of the energy gap. Upper right panel: optimal variational
	parameter $\delta$ for periodic boundary conditions. Lower left panel:
	energy per bond calculated with ED (circles) and VMC (squares) on 15 sites
	with open boundary conditions. Note that the optimal variational parameter
	$\delta=0$ in that case. Lower right panel: energy per bond calculated
	with VMC with periodic boundary conditions.	
	}
	\label{rlg-N6m2-gap-delta-nnc}
\end{figure}

This case is a priori more challenging to study because the relevant operator
that is generated in the renormalization group theory appears at higher order
than the one-loop approximation. Therefore, the gap can be expected to be
significantly smaller than in the previous case. This trend is definitely
confirmed by ED performed on clusters with up to 15 sites: the gap decreases
quite steeply with the system size (see upper left panel of
\gref{rlg-N6m2-gap-delta-nnc}). It scales smoothly however, and a linear
extrapolation points to a gap of the order $\Delta E\simeq 0.2$, much smaller
than in the \SU{6} case with $m=3$ ($\Delta E\simeq 4$), but finite. On the
largest available cluster, the bond energy has a significant dependence on the
bond position, with an alternance of two very negative bonds with a less
negative one.

These trends are confirmed and amplified by VMC. Indeed, the trimerized wave
function $\gwf{3}{\delta}$ leads to a better energy for all sizes, and the
optimal value scales very smoothly to a small but finite value
$\deltaopt\approx 0.03$. This value is about an order of magnitude smaller
than in the \SU{6} case with $m=3$, but the fact that it does not change with
the size beyond 60 sites is a very strong indication that the system
trimerizes (by contrast to the marginal case shown in \gref{rlg-E-delta}). The
trimerization is confirmed by the lower plots. For $n=15$, the VMC results are
again in nearly perfect agreement with ED, and for $n=60$, the bond energy
shows a very clear trimerization.

\begin{figure}[bh!]
	\begin{center}
		\includegraphics[scale=0.7]{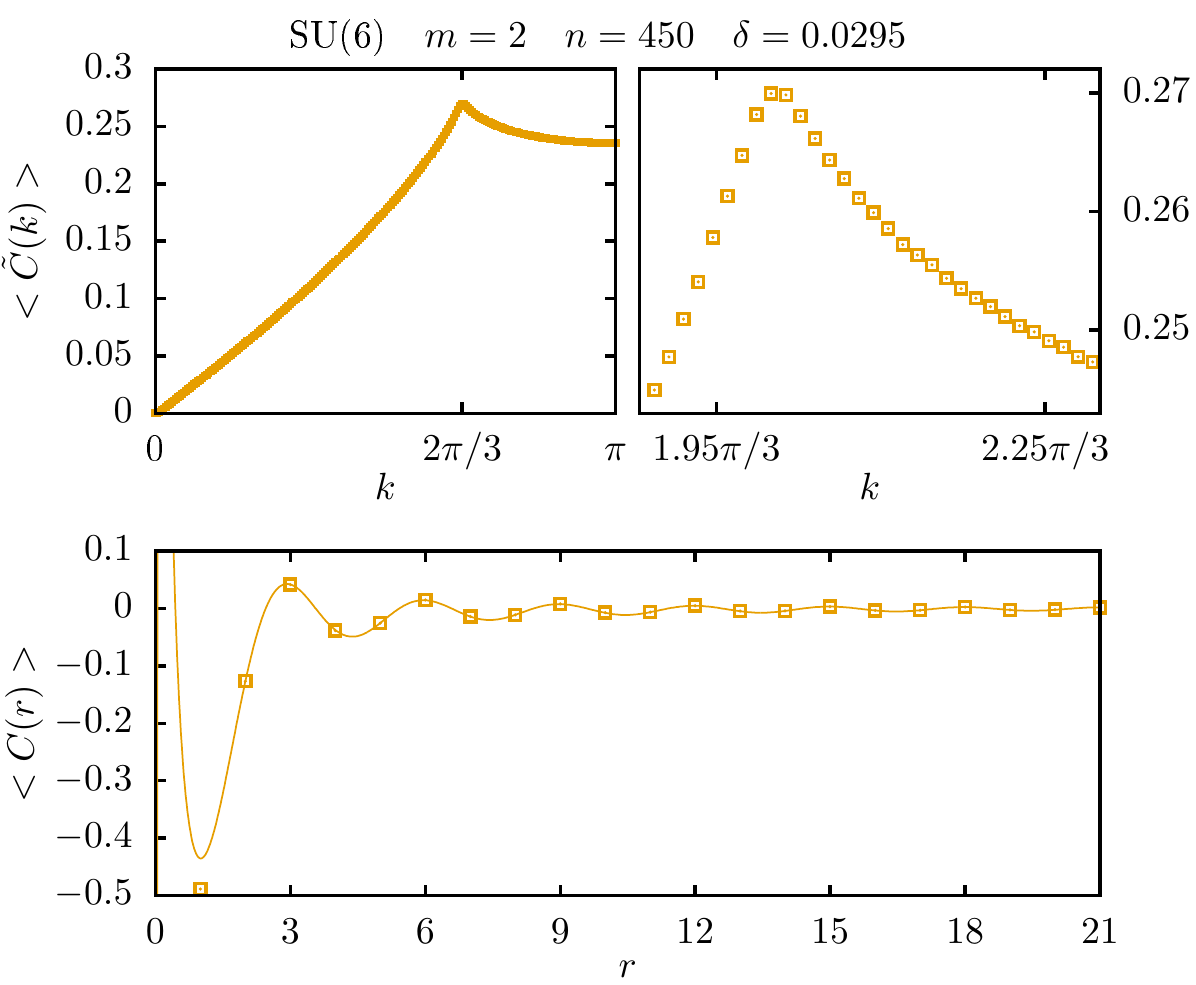}
	\end{center}
	\vspace{-0.6cm}
	\caption{
	Upper left panel: Structure factor of the \SU{6} model with $m=2$
	calculated with VMC using a trimerized wave function with the optimal
	variational parameter. Upper right panel: zoom on the region near
	$k=2\pi/3$. It clearly shows that the structure factor is smooth. Lower
	panel: real-space diagonal correlations for 60 sites.
	}
	\label{rlg-N6m2-sf-lrc}
\end{figure}

To test the nature of the long-range correlations is of course more challenging
than in the previous case since a small gap implies a long correlation length.
And indeed, on small to intermediate sizes, the structure factor has a sharp
peak at $k=2\pi/3$ very similar to the \SU{3}, $m=1$ case.  However, going to
very large system sizes (up to $n=450$ sites), it is clear that the concavity
changes sign upon approaching $k=2\pi/3$ (see upper right panel of
\gref{rlg-N6m2-sf-lrc}), consistent with a smooth peak, hence with
exponentially decaying correlation functions (see also lower panel of
\gref{rlg-N6m2-sf-lrc}).

In that case, in view of the small magnitude of the gap, hence of the very
large value of the correlation length, it would be difficult to conclude that
the system is definitely trimerized on the basis of ED only. In that respect,
the VMC results are very useful. On small clusters, the Gutzwiller projected
wave function with trimerization is nearly exact, and VMC simulations on very
large systems strongly support the presence of a trimerization and of
exponentially decaying correlations\footnote{We have been informed by S.
Capponi that these conclusions agree with unpublished DMRG results (S. Capponi,
private communication)}.

\section{Marginal cases: \SU{8} with $m=2$ and \SU{9} with $m=3$}

These two systems are the only ones which possess operators with scaling
dimension $\chi=2$. They are therefore the only cases where it is impossible to
predict whether the system is algebraic or gapped on the basis of Affleck's
analysis. As far as numerics is concerned, these cases can again be expected to
require large system sizes to conclude.

\begin{figure}[h]
	\begin{center}
		\includegraphics[scale=0.7]{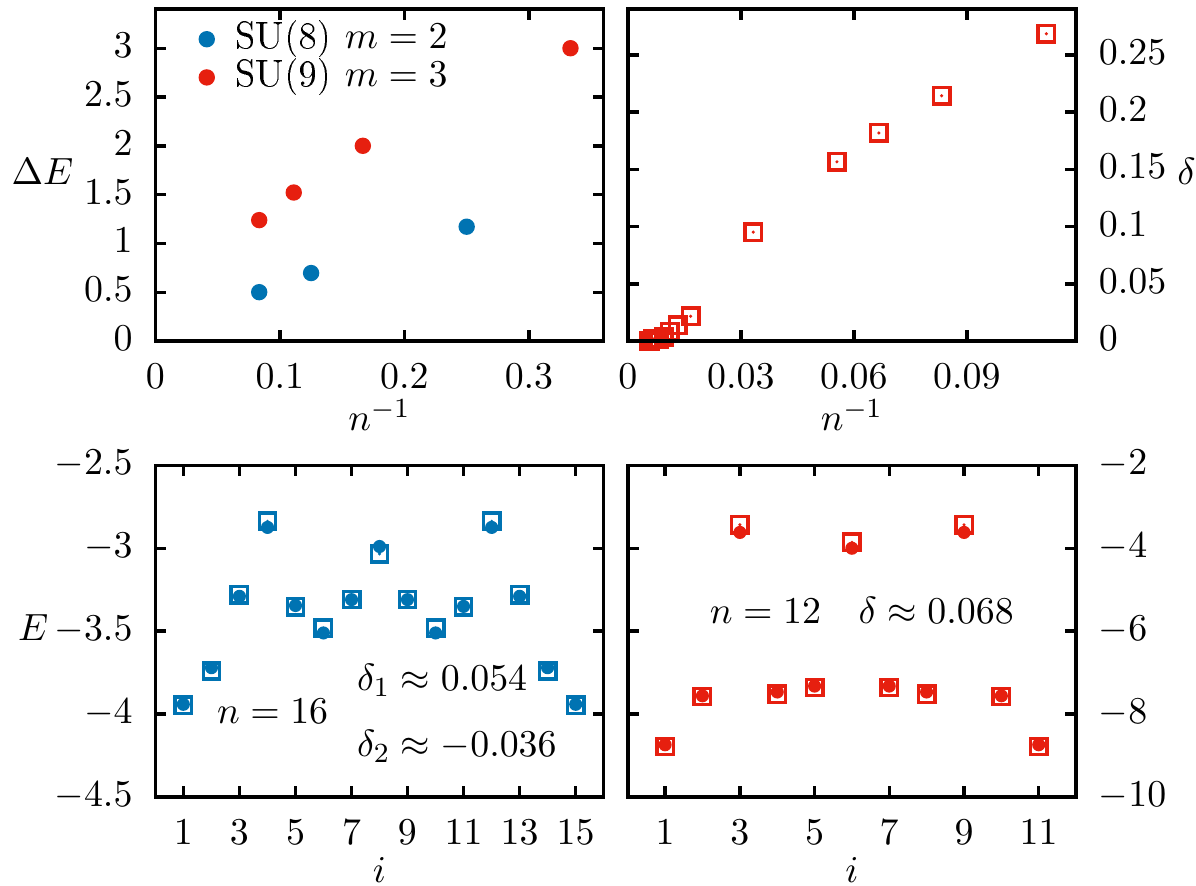}
	\end{center}
	\vspace{-0.5cm}
	\caption{
	ED and VMC results for the marginal cases \SU{8} with $m=2$ and \SU{9}
	with $m=3$. Upper left panel: size dependence of the energy gap for both
	cases. Upper right panel: optimal variational parameter $\delta$ for the
	\SU{9} case with periodic boundary conditions. The results for \SU{8} are
	not shown because they identically vanish for periodic boundary
	conditions. Lower left panel: energy per bond for \SU{8} calculated with
	ED (circles) and VMC (squares) for open boundary conditions. Note that the
	optimal variational parameters $\deltaopt$ are different from zero with
	open boundary conditions. Lower right panel: energy per bond for \SU{9}
	calculated with ED (circles) and VMC (squares) for open boundary
	conditions.	
	}
	\label{rlg-chieq2-gap-delta-nnc}
\end{figure}

The ED results are quite similar to the previous case. The scaling of the gap
is less conclusive because the last three points build a curve that is still
concave and not linear like in the previous case (see the upper right panel of
\gref{rlg-chieq2-gap-delta-nnc}). So one can only conclude that if there is a
gap, it is very small, especially for \SU{8} with $m=2$. The bond energies
build a pattern which is consistent with a weak tetramerization for \SU{8} with
$m=2$, and with a significant trimerization comparable to the \SU{6}, $m=2$
case for \SU{9} $m=3$.

\begin{figure}[h]
	\begin{center}
		\includegraphics[scale=0.7]{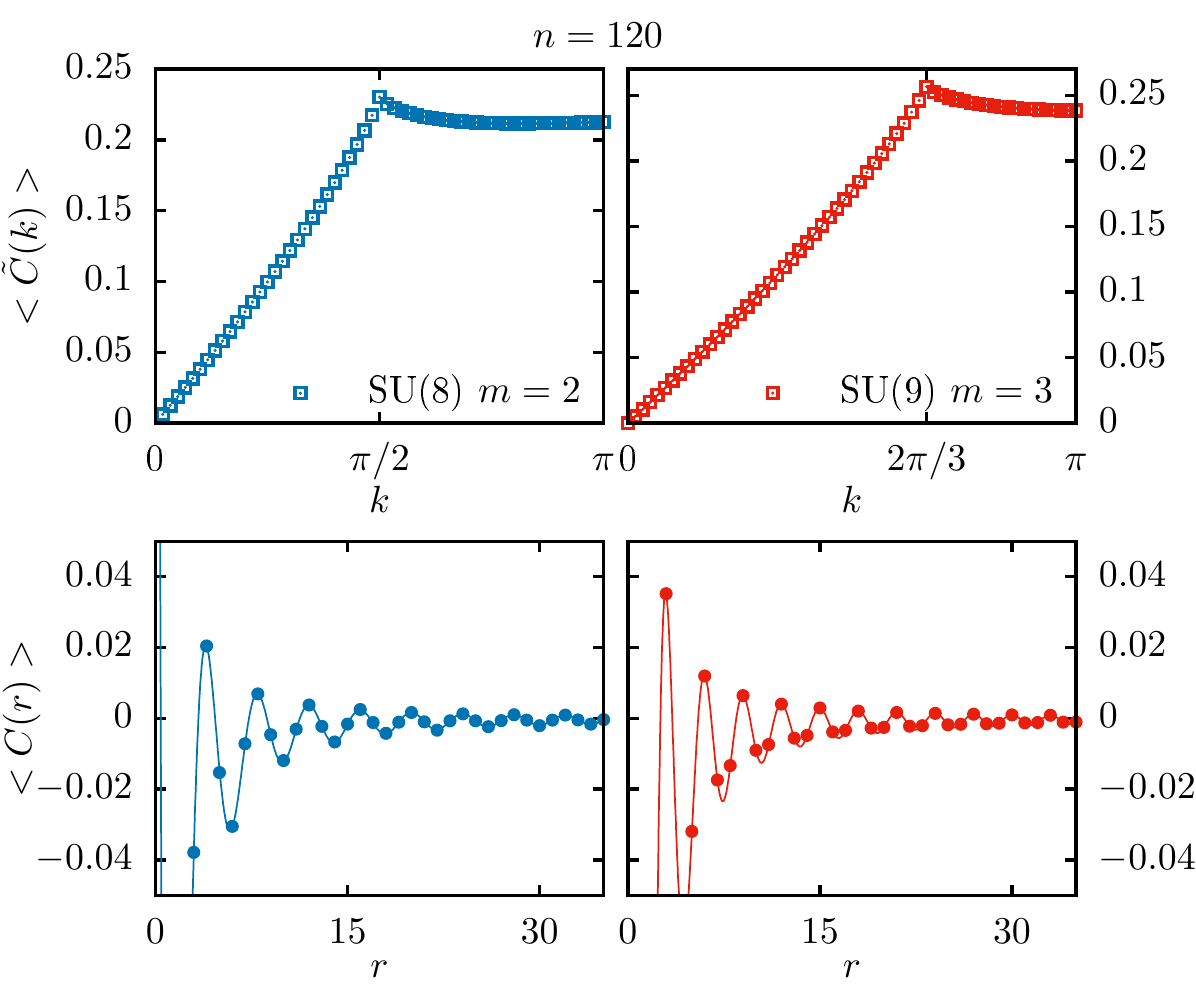}
	\end{center}
	\vspace{-0.5cm}
	\caption{
	Upper panels: Structure factor of the \SU{8} model with $m=2$ (left) and
	of the \SU{9} model with $m=3$ (right) calculated with VMC with periodic
	boundary conditions. Lower panels: real space correlations. The four plots
	represents results obtained with $\fwf$.
	}
	\label{rlg-chieq2-sf-lrc}
\end{figure}

The VMC method turns out to give a rather different picture however. For
\SU{8} with $m=2$, two variational wave functions ($\fwf,\gwf{4}{\delta_{1},
\delta_{2}}$) can be tested. Interestingly, for $n=16$ with open boundary
conditions, $\fwf$ fails to reproduce the bound energies pattern observed with
ED but $\gwf{4}{0.054,-0.036}$ is successful (see lower left pannel of
\gref{rlg-chieq2-gap-delta-nnc}). This pattern, which could be interpreted as
a weak tetramerization, is in fact probably just a consequence of the
four-fold periodicity of algebraic correlations in the presence of open
boundary conditions. Indeed, it turns out that, for any system size with
periodic boundary conditions, the minimization of the energy using
$\gwf{4}{\delta_{1},\delta_{2}}$ failed to find a solution for any
$\left\vert\delta_{1,2}\right\vert>0.002$. Therefore $\fwf$ is believed to be
the best variational wavefunction. The conclusion is that there is no
tetramerization, and that the correlations must be algebraic. This is also
supported by the structure factor, which seems to have a singularity at
$k=\pi/2$ (see \gref{rlg-chieq2-sf-lrc}).

\begin{figure}[h]
	\begin{center}
		\includegraphics[scale=0.7]{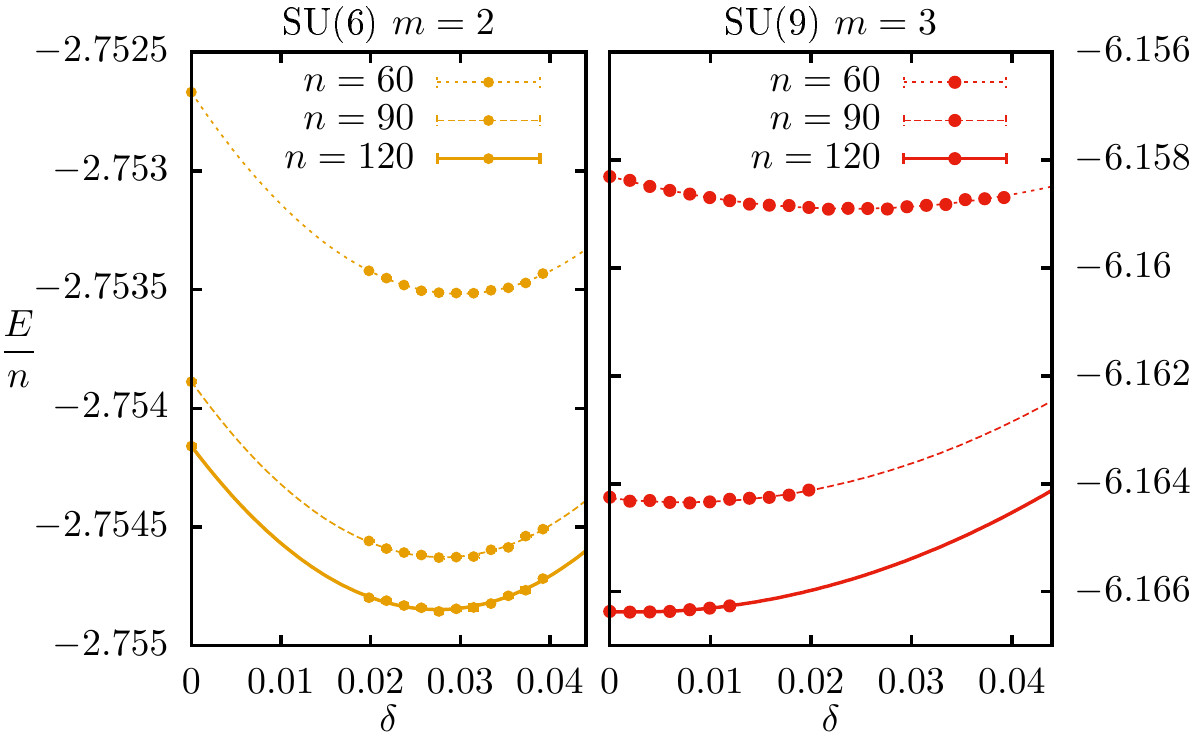}
	\end{center}
	\vspace{-0.5cm}
	\caption{
	Energy per site as a function of the variational parameter $\delta$ for
	the \SU{6} with $m=2$ (left) and \SU{9} with $m=3$ (right).
	}
	\label{rlg-E-delta}
\end{figure}

Let us now turn to \SU{9} with $m=3$. This system could in principle be
trimerized, and therefore $\fwf$ and $\gwf{3}{\delta}$ have been compared. For
small clusters, there is a large optimal value of $\delta$, actually much
larger than for \SU{6} with $m=2$, and the bond energies are typical of a
strongly trimerized system, in agreement with ED. However, $\deltaopt$
decreases very fast with $n$ until it vanishes for $n\gtrsim 100$ whereas, for
\SU{6} with $m=2$, $\deltaopt$ levels off at a finite value beyond $n=60$ (see
\gref{rlg-E-delta}). We interpret this behavior as indicating the presence of
a cross-over: on small length scales, the system is effectively trimerized,
but this is only a short-range effect, and the system is in fact gapless with,
at long-length scale, algebraic correlations.

One can again calculate the structure factor using the best variational wave
function (in both cases $\fwf$ for big enough systems) to check if a
discontinuity exists. The results displayed in the upper plots of
\gref{rlg-chieq2-sf-lrc} clearly show a discontinuity at $k=\pi/2$ for the
\SU{8} and at $k=2\pi/3$ for \SU{9}. These discontinuities indicate an
algebraic decay of the long-range correlations. The lower plot shows that even
if these systems are gapless, there is a maxima of the correlation every $N/m$
sites.

\section{Example with irrelevant operator: \SU{10} with $m=2$}

For completeness, we have also looked at a case where there is an irrelevant
operator of scaling dimension larger than 2, namely \SU{10} with $m=2$. As
expected, the best variational wave function is $\fwf$ for all sizes, and the
structure factor exhibits discontinuities at $k=2\pi/5$, consistent with a
gapless spectrum and algebraic correlations.

\section{Critical exponents}\label{rls-exponents}

Motivated by the remarkably accurate results obtained in previous works for the 
case the fundamental representation\cite{paramekanti_2007,wang_z2_2009}, we have
tried to use the VMC results to determine the critical exponent that controls
the decay of the correlation function at long distance, \eref{rle-sf}. For the
particular case of gapless systems, conformal field theory predicts an
algebraic decay of the long-range correlations function according to:
\begin{equation*}
	C\ep{r} = \dfrac{c_{0}}{r^{2}}+\dfrac{c_{k}\cos{2\pi r m/N}}{r^{\eta}}
\end{equation*}
where $\eta=2-2/N$ is the critical exponent. 

For systems with periodic boundary conditions, one can define two distances between
two points, which naturally leads to the following fitting function~\cite{Frischmuth1999}:
\begin{equation*}
	c_{0}(r^{-\nu}+\ep{n-r}^{-\nu})+c_{k}\cos{2\pi r m/N}(r^{-\eta}+\ep{n-r}^{-\eta})
\end{equation*}
with four free parameters: $c_0$, $\nu$ and $c_k$, $\eta$,  the amplitudes and
critical exponents of the components at $k=0$ and $k=2\pi m/N$ respectively.

There is a large degree of freedom in the choice of the fitting range. One
could in principle select any arbitrary range of sites $\ec{x_{i},x_{f}}$,
$0\leq x_{i}<x_{f}\leq n-1$. The problem is that each range will give different
critical exponents. In order to obtain some meaningful results the following
method has been chosen. Using the periodicity of the systems, only the ranges
with $x_{i}=a$ and $x_{f}=n-a-1$, $1\leq a\leq n/2$, have been considered. For
each value of $a$, the coefficient of determination of the fit has been
computed and if its value is higher than $0.999$ than the range $\ec{a,n-a-1}$
is selected to perform the extrapolation of the critical exponents. If the
value is too low, the fit is considered to be bad and the range with
$a\leftarrow a+1$ is tested. If no good range can be found with this criterion,
the condition over the coefficient of determination is relaxed to be higher
than $0.995$ and the first fit with a residual sum of squares divided by $n$
that is smaller than $10^{-7}$ is selected.

\begin{figure}[h]
	\begin{center}
		\includegraphics[scale=0.7]{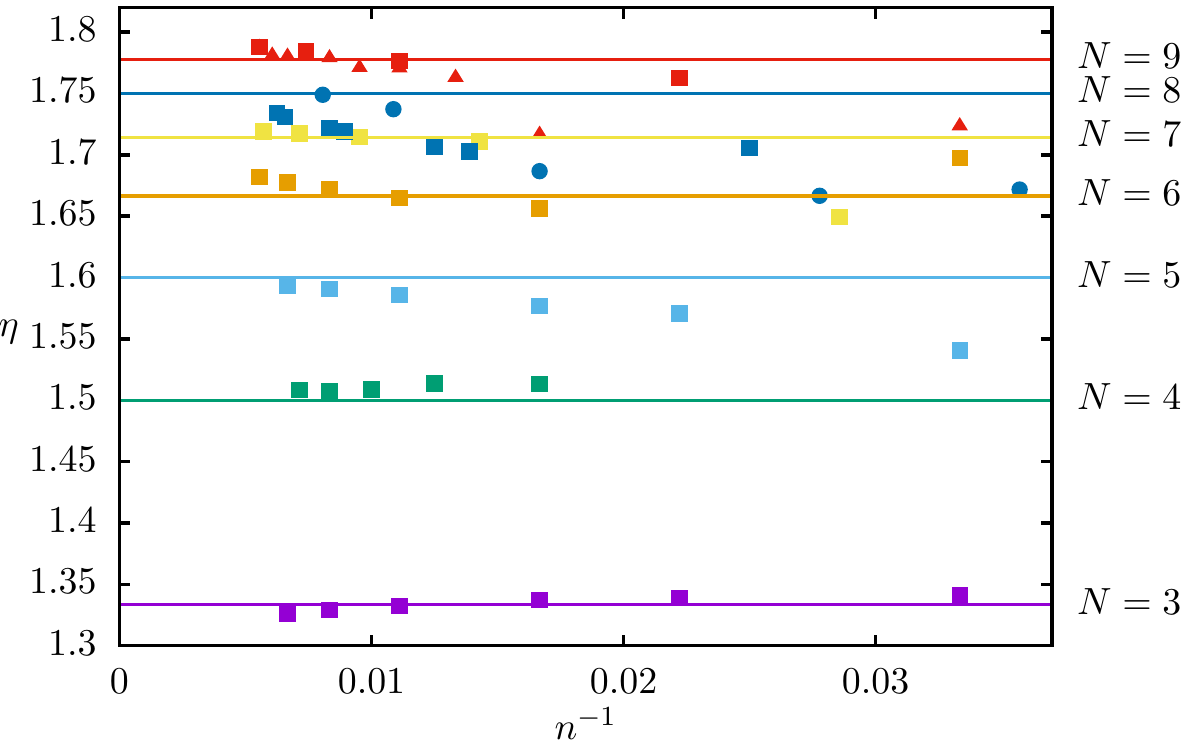}
	\end{center}
	\vspace{-0.5cm}
	\caption{
	Critical exponents $\eta$ of the gapless systems as a function of the
	system size. The squares, circles and triangles correspond respectively to
	$m=1,2$, and $3$ particles per site. All values given here have been
	calculated with $\fwf$.
	}
	\label{rlg-critical-exponents}
\end{figure}

The critical exponents $\eta$ obtained in this way are shown in
\gref{rlg-critical-exponents}.  The theoretical values of the critical
exponents $\eta = 2-2/N$ are shown as straight lines. In all cases, the
extracted exponents agree quite well with the theoretical predictions when $n$
is large enough. In particular, for a given $N$, the exponent $\eta$ does not
depend on $m$, as predicted by non-abelian bosonization. The critical exponents
$\nu$ has also been extracted but, as already observed~\cite{Frischmuth1999}, a
precise estimate is difficult to get. Nevertheless, for $N=3,4$,
$\nu\in\ec{1.8,2.25}$ and for $N\geq 5$, $\nu\in\ec{1.95,2.05}$ for the largest
systems.

\section{Conclusions}

Using variational Monte Carlo based on Gutzwiller projected wave functions, we
have explored the properties of \SU{N} Heisenberg chains with various totally
antisymmetric irreps at each site. In the case of the fundamental
representation, which is completely understood thanks to Bethe ansatz and to
QMC simulations, these wave functions are remarkably accurate both regarding
the energy and the long-range correlations. In the case of higher
antisymmetric irreps, where field theory arguments are in most cases able to
predict that the system should be gapless or gapped, allowing for a symmetry
breaking term in the tight binding Hamiltonian used to define the unprojected
wave function leads to results in perfect agreement with these predictions,
and the ground state is found to be spontaneously dimerized or trimerized.
Finally, in the two cases where the operator that could open a gap is
marginal, \SU{8} with $m=2$ and \SU{9} with $m=3$, this variational approach
predicts that there is no spontaneous symmetry breaking, and that correlations
decay algebraically. These results suggest that the operators are marginally
irrelevant in both cases. It would be interesting to test these predictions
either analytically by pushing the renormalization group calculations to
higher order, or numerically with alternative approaches such as DMRG or QMC.

In any case, these results prove that Gutzwiller projected fermionic wave
functions do a remarkably good job at capturing quantum fluctuations in
one-dimensional \SU{N} Heisenberg models with totally antisymmetric irreps.
Considering the encouraging results obtained in 2D for the \SU{N} Heisenberg
model with the fundamental irrep at each site on several lattices, one can
legitimately hope these wave functions to be also good for the \SU{N}
Heisenberg model with totally antisymmetric irreps at each site in 2D. Work is
in progress along these lines.


We acknowledge useful discussions with S. Capponi, M. Lajko, P. Lecheminant,
L. Messio, and K. Penc. This work has been supported by the Swiss National
Science Foundation.

\bibliography{vmc-sun-heisenberg-chain-BIB}
\end{document}